\documentclass[a4paper,aps,prl,twocolumn,floatfix,showpacs,superscriptaddress]{revtex4-1}
\usepackage[utf8]{inputenc}
\usepackage{graphics}
\usepackage{epsfig}
\usepackage{times}
\usepackage{xcolor}   
\usepackage{amsfonts}
\usepackage{amssymb}
\usepackage{amsthm}
\usepackage{bm} 
\usepackage{amsmath}
\usepackage{xr-hyper} 
\usepackage{hyperref}
\usepackage{color}
\usepackage{graphicx}
\usepackage{subfigure}
\usepackage{textcomp}
\usepackage{wasysym}
\usepackage{soul}
\usepackage{verbatim}
\usepackage[normalem]{ulem}

\def\coup{J} 
\def\spin{\mathbf{S}} 

\begin{document}

\title{Topology and geometry of spin origami}

\author{Krishanu Roychowdhury}
\affiliation{Laboratory of Atomic And Solid State Physics, Cornell University, Ithaca, NY 14853.}
\affiliation{Kavli Institute for Theoretical Physics, University of California, Santa Barbara, CA 93106-4030.}
\author{D. Zeb Rocklin}
\affiliation{Laboratory of Atomic And Solid State Physics, Cornell University, Ithaca, NY 14853.}
\affiliation{School of Physics, Georgia Institute of Technology, Atlanta, GA 30332.}
\author{Michael J. Lawler\footnote{Corresponding author.}}
\affiliation{Laboratory of Atomic And Solid State Physics, Cornell University, Ithaca, NY 14853.}
\affiliation{Kavli Institute for Theoretical Physics, University of California, Santa Barbara, CA 93106-4030.}
\affiliation{Department of Physics, Binghamton University, Binghamton, NY, 13902.}

\begin{abstract}
Kagome antiferromagnets are known to be highly frustrated and degenerate when they possess simple, isotropic interactions.
 We consider the entire class of these magnets when their interactions are spatially anisotropic.
We do so by identifying a certain class of systems whose degenerate ground states can be mapped onto the folding motions of a generalized ``spin origami'' two-dimensional mechanical sheet.
Some such anisotropic spin systems, including ${\rm Cs}_2{\rm ZrCu}_3{\rm F}_{12}$, map onto flat origami sheets, possessing extensive degeneracy similar to isotropic systems. Others, such as ${\rm Cs}_2{\rm CeCu}_3{\rm F}_{12}$, can be mapped onto sheets with non-zero Gaussian curvature, leading to more mechanically stable corrugated surfaces. 
Remarkably, even such distortions do not always lift the entire degeneracy, instead permitting a large but sub-extensive space of zero-energy modes.
We show that for ${\rm Cs}_2{\rm CeCu}_3{\rm F}_{12}$, due to an additional point group symmetry associated with structure, these modes are 'Dirac' line nodes with a double degeneracy protected by a topological invariant. 
The existence of mechanical analogs thus serves to identify and explicate the robust degeneracy of the spin systems. 
\end{abstract}

\maketitle

\noindent
\label{intro}
Frustrated condensed matter such as kagome Heisenberg antiferromagnets (KHAF) possesses many degenerate ground states that can be either delicate or robust, despite being accidental in the sense of not being protected by a symmetry.
Isotropic KHAF have been mapped onto triangulated sheets of ``spin origami''~\cite{shender1993kagome,chandra1993anisotropic, Ritchey1993spin},
revealing that, at the classical level, these materials can have as many ground states as there are ways to fold a sheet of paper with one crease for each atomic spin. 
Splitting this degeneracy by making the magnetic moments spin-$1/2$ would permit the formation of a quantum spin liquid\cite{yan2011spin}, 
but ``clearly the KHAF is a problem where competing states of very different character lie very close in energy''~\cite{liao2017gapless}. Like many other strongly correlated materials, a complex phase diagram arises and to our knowledge no general explanation has even been proposed. However, at least in the classical large-$S$ limit, it appears that recent advances in the study of metamaterials~\cite{sun2012surface, babaee20133d,  schenk2013geometry, wei2013geometric, kane2014topological, shan2014harnessing, paulose2015selective, rocklin2015transformable, lubensky2015phonons, rocklin2016mechanical, abbaszadeh2016sonic, chen2016topological}, such as origami, suggest just such an explanation. 

Mechanical systems are among the oldest subject to formal study, yet today mechanical metamaterials display new properties and states of matter derived purely from their structure. Many such systems rely on a counting argument developed by Maxwell to determine mechanical stability by counting degrees of freedom (d.o.f.) and constraints~\cite{maxwell1864xlv} and extended by Calladine to account for redundant constraints~\cite{calladine1978buckminster}.
Recently, Kane and Lubensky~\cite{kane2014topological} relied on this count to discover, in the context of ball and spring systems, that systems could display exotic zero-energy boundary modes when they had equal numbers of d.o.f. and constraints. In an initially gapped system the difference between these quantities, labeled $\nu$, can only go from 0 to 1, indicating the appearance of a zero mode, when the gap closes. In this context, called ``isostatic'', $\nu$ itself is a topological invariant. Further, they build a local version of Maxwell counting and derive a winding number topological invariant for phonon band structures which demands edge states in ``polarized'' isostatic systems~\cite{kane2014topological}, bulk solitons in isostatic one dimensional systems~\cite{chen2014nonlinear}, and Weyl point nodes in isostatic two dimensional systems~\cite{po2016phonon, rocklin2016mechanical}. In systems with translational symmetries, such a gap trivially closes at wavevector $\mathbf{k} = 0$, but survives for spatially varying modes. Thus, by combining energy gaps with Maxwell counting, a topological mechanics emerges that connects zero modes to topological invariants.

This discovery brings new meaning to Moessner and Chalker's two seminal papers~\cite{moessner1998properties,moessner1998low} that exploited Maxwell counting to shed light on the accidental ground state degeneracy of classical kagome and a few other antiferromagnets. Grouping the terms in the Hamiltonian into constraints, a procedure that underlies the spin origami construction, they argue Maxwell's $\nu$ is often a useful measure of frustration in frustrated magnets. They show that $\nu > 0$ in the pyrochlore Heisenberg antiferromagnet and demands zero modes while $\nu$ vanishes in the isotropic kagome KHAF so that its zero modes must arise from a redundancy among the constraints. This redundancy renders the kagome case complex from this perspective, but since it has $\nu=0$, like Kane and Lubensky's isostatic systems, this complexity should come with topological invariants that could provide an alternative explanation of kagome zero modes. 

In this Letter, guided by the concepts of topological mechanics, we study how topology and geometry explicate magnetic frustration in kagome antiferromagnets. Specifically, we solve for the ground states of a class of distorted KHAF obeying a condition (necessary and sufficient) under which the ground states of those systems possess origami analogs. We further identify ${\rm Cs}_2{\rm ZrCu}_3{\rm F}_{12}$ and ${\rm Cs}_2{\rm CeCu}_3{\rm F}_{12}$ as candidate materials that can foster such a spin origami state. Surprisingly, the origami we predict for ${\rm Cs}_2{\rm ZrCu}_3{\rm F}_{12}$ is flattenable like the original spin origami construction of isotropic kagome antiferromagnets despite possessing spatial anisotropies in the spin exchanges. It thus also features a flat band in its spin wave dispersions. In distinction, the origami we find for ${\rm Cs}_2{\rm CeCu}_3{\rm F}_{12}$ is nonflattenable and mechanically more rigid. Nevertheless it retains a finite residual entropy that has dramatic consequences -- doubly degenerate topological ``Dirac'' lines nodes in the spin wave dispersions akin to the Fermi surface of a metal. We discover these lines of zero modes follow from a combination of a special point group symmetry of our predicted nonflattenable periodic origami and a $\mathbb{Z}_2$ topological invariant we build from this symmetry and its isostatic property. In passing, we also find singly degenerate topological ``Weyl'' lines of zero modes follow from a similar $\mathbb{Z}_2$ topological invariant for generic periodic origami due to their mysterious realness property~\cite{chen2016topological}. Thus, we show these ``origami magnets'' have robust accidental degeneracy by applying recent developments in the study of metamaterials to that of kagome antiferromagnets.  

\begin{figure}
\centering
 \includegraphics[width=7.0cm]{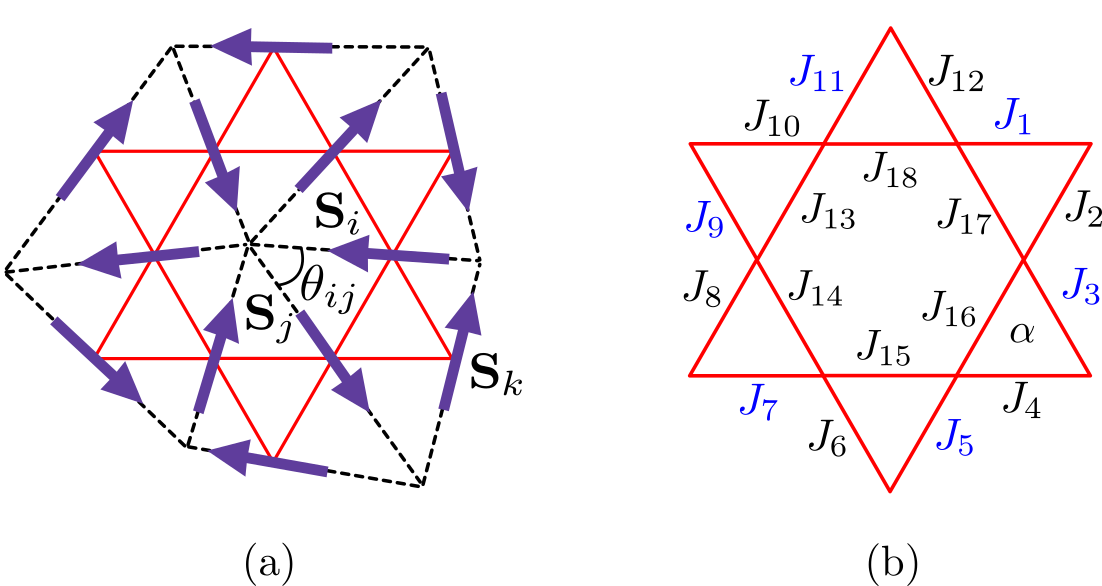}
 \caption{(a) Mapping from a spin configuration on the star of David to an origami where the spins in the former represent the edge vectors in the latter drawn in dotted lines. (b) A kagome ``Star of David'' with nonuniform interactions which on the exterior bonds satisfy the star condition (Eq.~\ref{eq:star}), necessary for a generic spin system to possess an origami analog. The expression of the interior angle $\theta_{ij}$ is given in Eq.~(\ref{eq:angle1}).}
 \label{figg1}
\end{figure}

We define a generic KHAF by
\begin{equation}
 \mathcal{H} =  \sum_{\langle i,j \rangle,\alpha} \coup_{ij}{\bf S}_i\cdot {\bf S}_j = \frac{1}{2} \sum_{\triangle\alpha,\triangle'\beta} S_{\triangle\alpha} J^{\triangle\alpha,\triangle'\beta} S_{\triangle'\beta} + {\rm{const.}},
 \label{eq:ham1}
\end{equation}
where $\alpha\in\{x,y,z\}$ denote the spin components of the spin vector ${\bf S}_i$, $J^{\Delta\alpha,\Delta'\beta}$ is a positive definite symmetric matrix, $S_{\Delta\alpha} = \ell^\Delta_iS_{i\alpha} + \ell^\Delta_jS_{j\alpha}+\ell^\Delta_kS_{k\alpha}$ with $\Delta$ denoting a triangle with sites $ijk$, and $\ell^\Delta_i$ are (dimensionless) positive real numbers. This form can be worked out straightforwardly for exclusively nearest neighbor exchanges. The result is $J^{\Delta\alpha,\Delta'\beta} = J_\Delta \delta_{\Delta,\Delta'}\delta_{\alpha,\beta}$, $J_\Delta > 0$ and $\ell^\Delta_i = \sqrt{J_{ij}J_{ik}/J_\Delta J_{jk}}$ for triangle $\Delta =\langle ijk\rangle$. The zero-energy condition then requires that the fixed-length vectors $\ell^\Delta_i S_{i\alpha}$ on a triangle sum to zero ($S_{\Delta\alpha}=0$), the very condition that is met by vectors along the edges of a rigid triangle of the type shown in Fig.~\ref{figg1} (a), provided the anisotropy is not so strong that the triangle inequality $\ell^\Delta_i < \ell^\Delta_j + \ell^\Delta_k$ or its cyclic permutations are violated. For the case of isotropic KHAF, these triangles permitted the mapping of zero-energy configurations onto folding patterns of an origami sheet consisting of equilateral triangular faces~\cite{shender1993kagome,chandra1993anisotropic, Ritchey1993spin}.

For an inhomogeneous system, however, we cannot guarantee the existence of an origami analog merely by satisfying $\sum_\Delta \ell_i \mathbf{S}_i=0$. This mapping specifies the shape of the triangular face but not its scale; since each edge corresponds to two faces but can have only one length ($\ell^\Delta_i = \ell^{\Delta'}_i$), an additional requirement emerges on the couplings around a magnetic system such as those found in Fig.~\ref{figg1} (b) (see Supplementary Material):
\begin{align}
 J_1J_3J_5J_7J_9J_{11}=J_2J_4J_6J_8J_{10}J_{12},
 \label{eq:star}
\end{align}
where here we explicitly labeled the bonds of the lattice for clarity. As we will see, this condition is met for some but not all KHAF systems. It is a necessary and sufficient condition for the existence of a particular (up to overall scale) origami analog. However, even among such systems, an important distinction arises depending on the geometry of the origami.

Vertices satisfying Eq.~(\ref{eq:star}) are not in general flat. The interior angle of the triangular surface associated with, e.g., the triangle formed by $\spin_i$, $\spin_j$, and $\spin_k$ in Fig.~\ref{figg1} i.e. the angle between $\spin_i$ and $\spin_j$ is given by
\begin{align}
\label{eq:angle1}
\theta_{ij} = \cos^{-1}\bigg[\frac{1}{2} \bigg( \frac{J_{ik}}{J_{jk}} + \frac{J_{jk}}{J_{ik}} - \frac{J_{ik}J_{jk}}{J_{ij}^2} \bigg)\bigg].
\end{align}
We can compute them directly from the exchange constants. It is only the special case for which the sum over the angles about a vertex is $2\pi$ when the vertex can be formed from a flat sheet, the condition that is usually (but not always~\cite{xuehoutan2005discrete, santangelo2017extreme}) assumed for origami. ``Non-Euclidean'' vertices violate this, and are said to have nonzero discrete Gaussian curvature (they are nonflattenable, as described in the supplementary material) equal to the angle deficit~\cite{meyer2003discrete}
\begin{align}
\label{eq:angle2}
\mathcal{G}_{\hexagon} = 2\pi - \sum_{\langle ij \rangle \in \hexagon} \theta_{ij},
\end{align}
where $\langle ij \rangle \in \hexagon$ denotes all adjacent pairs of spins $\spin_i$ and $\spin_j$ that meet at the vertex at the center of the hexagon $\hexagon$.
When this angle deficit vanishes, the spins adjoining the vertex can be and are expected~\cite{chubukov1992order} to be coplanar. In this case, each vertex possesses a zero mode corresponding to rotating the spins (edges) out of plane. In contrast, nonzero angle deficits preclude these local zero modes and necessarily lift the extensive degeneracy. 
Thus, the sign of each vertex's angle deficit, $\mu_{\hexagon} \equiv \rm{sgn}(\mathcal{G}_{\hexagon})$ is a topological invariant, in that it can change only when zero modes appear.

Note that these angle deficits, like the angles themselves, depend only on the coupling constants [via Eq.~(\ref{eq:angle1})] and not on the spin orientations. In the language of differential geometry, this is Gauss's ``Theorema Egregium'', that the Gaussian curvature is intrinsic to the system and does not depend on changes to its configuration that are isometries (zero modes)~\cite{gauss2013general, tapp2016differential}. Thus, degeneracy is determined not by fluctuations or dynamics but is largely determined by hidden geometric constraints. While individual vertices are governed by \emph{geometry}, they are collectively constrained to have zero total angle deficit, due to \emph{topological} constraints on the curvature given by the Gauss-Bonnet theorem as described in the supplementary material.

\begin{figure}
 \centering
 \includegraphics[width=8.0cm]{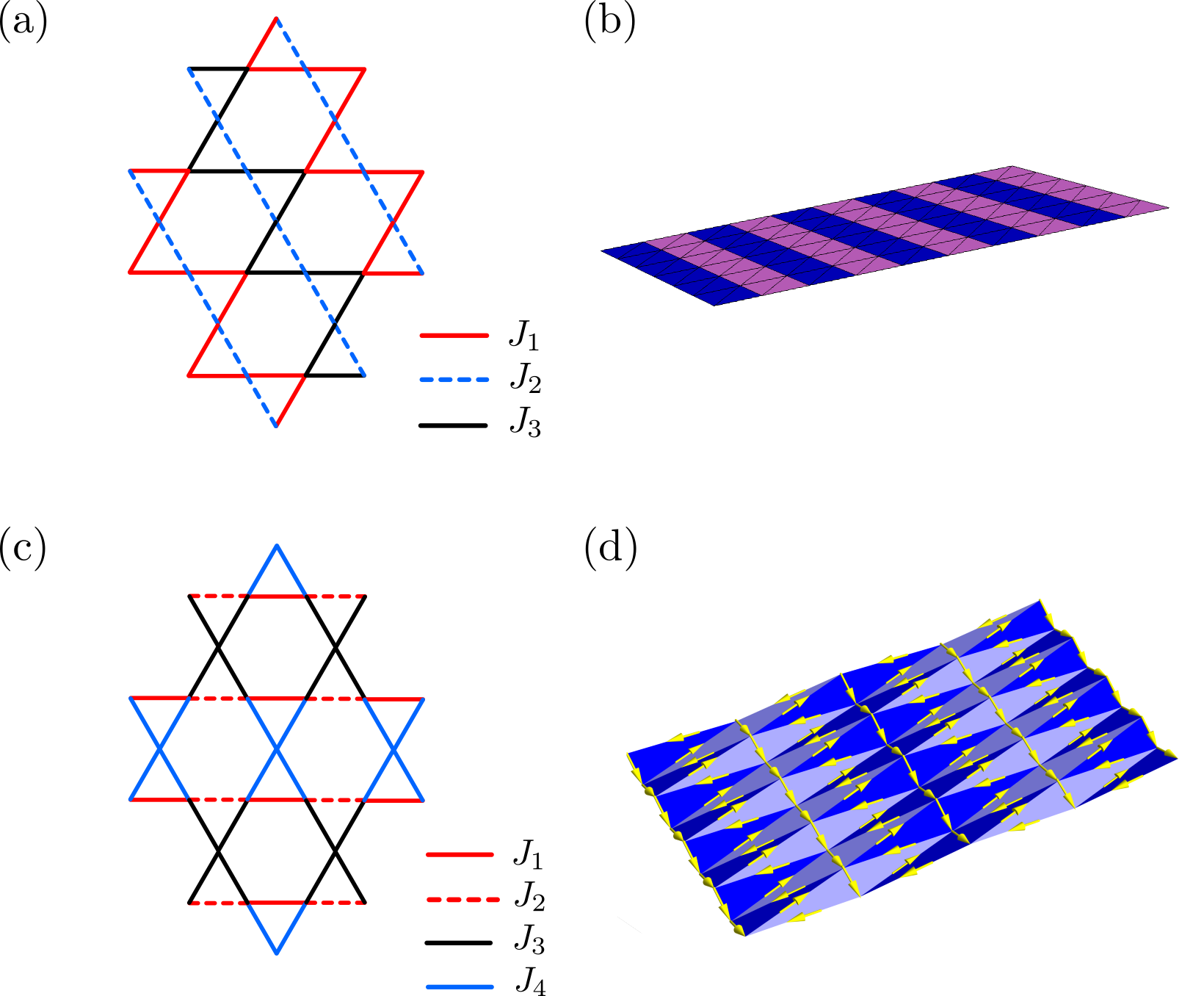}
 \caption{(a) The distorted kagome lattice structure of ${\rm Cs}_2{\rm ZrCu}_3{\rm F}_{12}$ with interactions that satisfy the star condition. (b) The origami analog of the $q=0$ state of (a) is a flat sheet consisting of isosceles triangles (spin arrows not shown). The dark blue and the purple faces correspond respectively to the blue-black and red-blue triangles of the kagome lattice in (a).
 (c) The distorted kagome lattice structure of ${\rm Cs}_2{\rm CeCu}_3{\rm F}_{12}$ with interactions obeying the star condition. (d) The spin origami for a $q=0$ state of (c) is a nonflattenable surface with coplanar pairs of triangles that form diamond shapes.}
\label{figg2}
\end{figure}

Among kagome materials that meet the star condition [Eq.~(\ref{eq:star})] despite distortion, we identify two that exemplify sharply distinct degeneracy. ${\rm Cs}_2{\rm ZrCu}_3{\rm F}_{12}$~\cite{ono2009magnetic, downie2014structural} has a pattern of spins shown in Fig.~\ref{figg2} (a) and (b) that, despite distortion, nevertheless lead to flat vertices. Hence, they resemble the isotropic spin origami previously studied~\cite{shender1993kagome,chandra1993anisotropic} despite their distortion. In contrast, ${\rm Cs}_2{\rm CeCu}_3{\rm F}_{12}$~\cite{amemiya2009partial}, as shown in Fig.~\ref{figg2} (c) and (d), 
with vertices having a finite curvature $\pm\mathcal{G}$ with
\begin{equation}
 \mathcal{G} = 4\cos^{-1}\frac{J_3}{2J_2} - 4\cos^{-1}\frac{J_4}{2J_1}.
\end{equation}
Evidently, $\mathcal{G}\leftrightarrow-\mathcal{G}$ when $J_{1,4}\leftrightarrow J_{2,3}$. The experimentally measured values of the interaction parameters $J_1=316$ K, $J_2=297$ K, $J_3=88$ K, and $J_4=85$ K (taken from Ref.~\cite{amemiya2009partial}) yields $\mathcal{G}\sim-0.055$. Straining the system tunes the interactions away from these values pushing the origami analog through a flat state and should therefore result in a topological phase transition in the sense of altering the invariant $\mu_{\hexagon}\equiv\rm{sgn}(\mathcal{G}_{\hexagon})$, as described in the supplementary material. Such a situation is experimentally conceivable as a controlled tuning of interactions in kagome systems has been achieved by means of applying pressure~\cite{wang2012pressure} or uniaxial stress~\cite{kuchler2017uniaxial}.

Given the ground state ordering patterns of the Fluoride materials shown in Fig.~\ref{figg2}, we now turn to the question of whether the associated spin waves in those materials have any special features. We can qualitatively understand the frustration associated with the zero modes of these two materials by borrowing the concept of self stresses from topological mechanics. In the mechanical analog of the flat spin origami sheet (as in ${\rm Cs}_2{\rm ZrCu}_3{\rm F}_{12}$), we can add tensions to the twelve edges of the six triangular faces adjoining a given vertex while preserving mechanical equilibrium regardless of the shapes of the coplanar faces. These self stress modes then imply the existence of zero modes since they correspond to redundancy of constraints functions in the triangle conditions~\cite{calladine1978buckminster}. These zero modes are displacements of vertices in the direction perpendicular to the faces. They are the manifestation in distorted kagome antiferromagnets with flattenable origami ground states of the zero modes existing in isotropic kagome antiferromagnets. However, for generic nonflattenable origami with non-coplanar edges, as in the ${\rm Cs}_2{\rm CeCu}_3{\rm F}_{12}$ compound, many of these self stresses are no longer possible---the rigidity of the sheet has become fundamentally enhanced via its geometry in a process akin to corrugation. This then has the effect of lifting the zero-energy band of phonons (lattice vibrations) from the origami system and magnons from the analogous spin system. 
The mechanical responses thus predict a flat band of spin waves associated with flattenable origami ground states (frustration preserved by distortions) but dispersing bands for nonflattenable origami and suggests $\mu$, mentioned above, may be a topological invariant whose change is associated with the emergence of a zero mode. So at this level we predict frustration can be relieved by the distortions in ${\rm Cs}_2{\rm CeCu}_3{\rm F}_{12}$ but not in ${\rm Cs}_2{\rm ZrCu}_3{\rm F}_{12}$.

\begin{figure}
\centering
 \includegraphics[width=8.6cm]{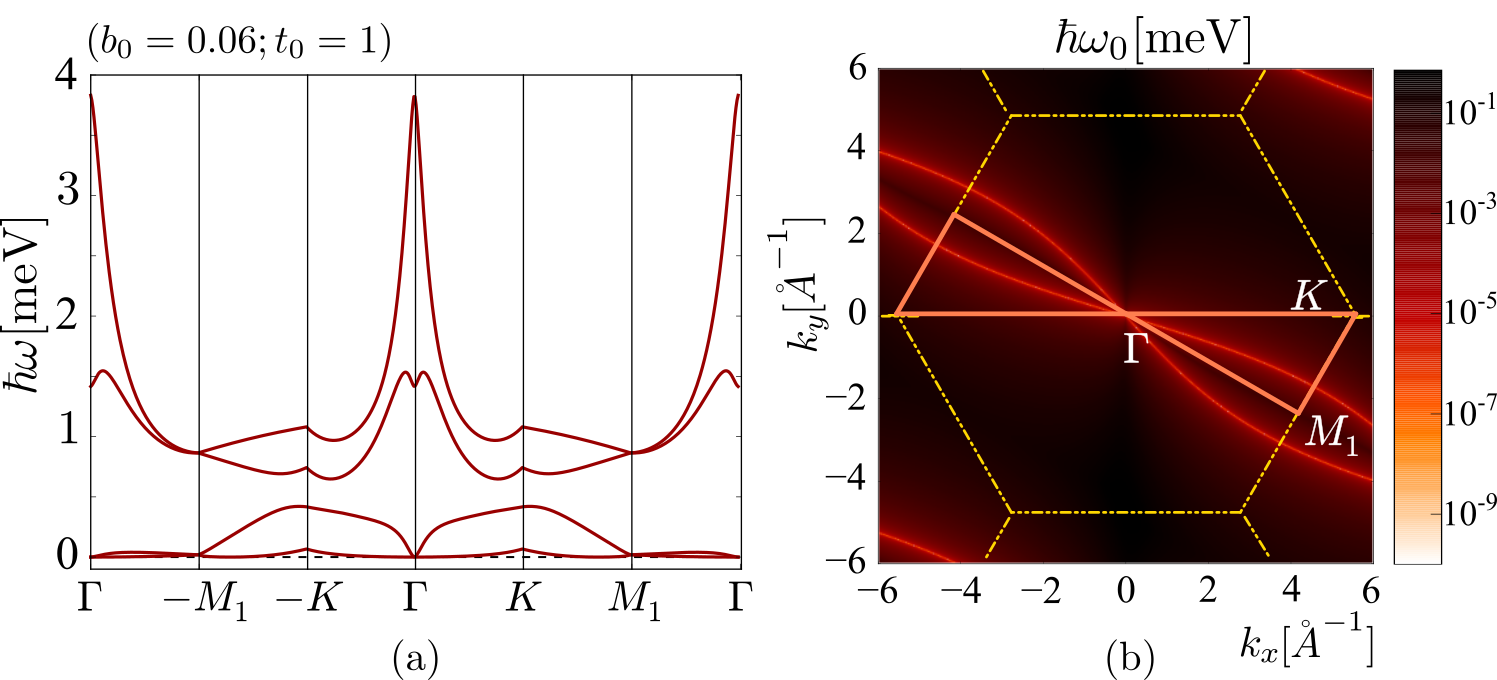}
 \caption{(a) Some of the lowest spin wave frequencies of ${\rm Cs}_2{\rm CeCu}_3{\rm F}_{12}$ as plotted along the high-symmetry path in the BZ shown in the inset and corresponding to the ground state specified by $b_0=0.06$ (see supplementary material for the definition of $b_0$). (b) A plot (in log scale) of the lowest frequency ($\omega_0$) in the BZ reveals the Dirac Line nodes.}
 \label{figg3}
\end{figure} 

\begin{figure*}
\centering
 \includegraphics[width=12.8cm]{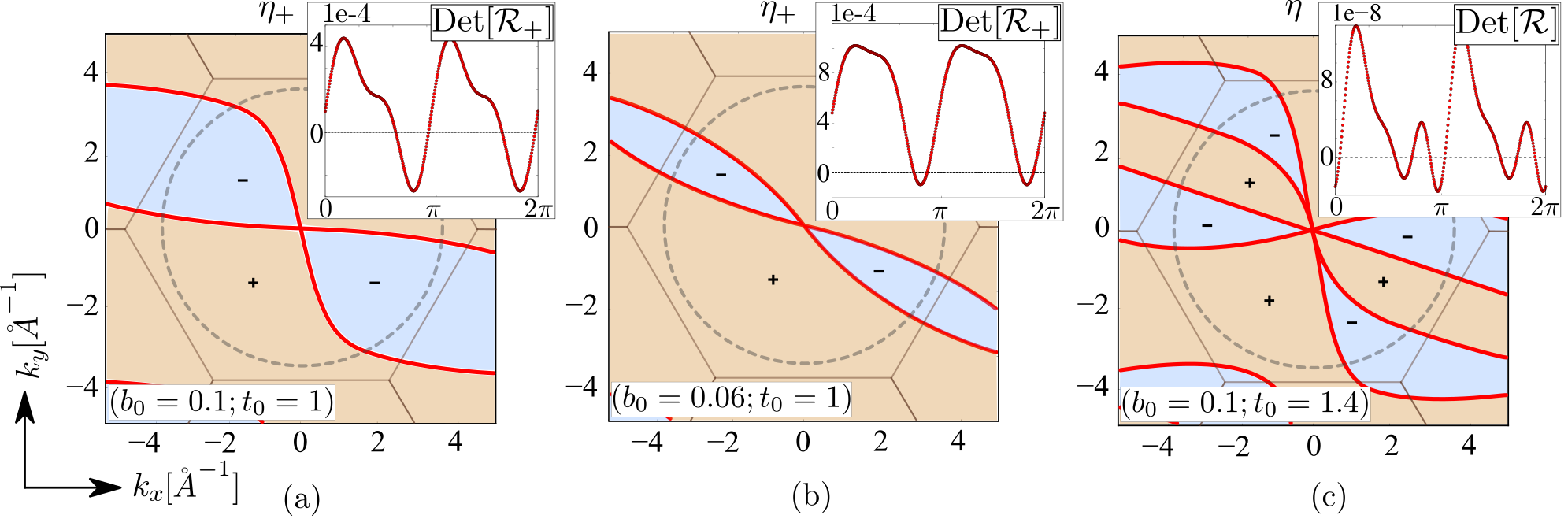}
 \caption{(a) and (b) Dirac line nodes (thick red lines) separating zones of different values of $\eta_+$ (yellow and blue correspond to `+' and `-' respectively) in the spin wave dispersions of ${\rm Cs}_2{\rm CeCu}_3{\rm F}_{12}$. We study these here for two different ground states (defined by the parameter $b_0$) that represent two members of the one dimensional family of origami configurations obtained for the periodic state (see supplementary material). The insets of (a) and (b) are the plots of $\eta_+$ over a circle in the BZ shown on the dotted line. The locations of the lines are decided by the condition $\text{Det}[\mathcal{R}({\bf{k}})]=0$ and depend on $b_0$. (c) Under deformations that break the point group symmetry of ${\rm Cs}_2{\rm CeCu}_3{\rm F}_{12}$, each Dirac line splits into two Weyl lines which are characterized by $\eta$ in Eq.~\ref{eq:topo}. The inset of (c) is the plot of $\eta$ over a circle in the BZ.}
 \label{figg4}
\end{figure*} 

We can learn more about the zero modes by considering the rigidity matrix~\cite{kane2014topological}. It characterizes the entire linear spin-wave theory of spin origami, which we choose to describe in terms of small spin rotations about the ground state using canonical variables $x^{i\mu} \equiv (q^i,p_i)$. From the constraint functions of the triangle condition, the rigidity matrix is just the leading term obtained by expanding in $x^{i\mu}$:
\begin{equation}
 \mathcal{R}_{\Delta\alpha,i\mu} = \frac{\partial S_{\Delta\alpha}}{\partial x^{i\mu}}.
 \label{eq:rig}
\end{equation}
The Hamiltonian matrix governing the spin waves is then $\mathcal{H}_{\rm SW} = \mathcal{R}^T\mathcal{R}$ where $\mathcal{R}$ is a square matrix because the number of constraints is equal to the number of degrees of freedom $\nu=D-K=0$. Solving for the spin wave frequencies, we find a flat band for the flattenable origami of ${\rm Cs}_2{\rm ZrCu}_3{\rm F}_{12}$ as expected but doubly degenerate ``Dirac'' line nodes for ${\rm Cs}_2{\rm CeCu}_3{\rm F}_{12}$ (see Fig.~\ref{figg3}). Existence similar line nodes have been previously reported in certain 3D topological semimetals (see Ref.~\cite{kim2015dirac} and references therein), however, not in magnetic systems or in 2D. So the rigidity matrix both encodes the flat spin wave band of a flat origami and reveals line nodes of nonflattenable origami. 

Zero modes occur precisely at those wavevectors for which $\text{Det}[ {\bf \mathcal{R}}({\bf k})]$ vanishes.This determinant is for general mechanical systems complex, leading to nonzero winding numbers
\begin{equation}
  w(C) = \frac{1}{2\pi} \oint_C d (\text{arg}\ \text{Det}[ {\bf \mathcal{R}}({\bf k})]),
  \label{eq:wind}
\end{equation}
around paths $C$ in the Brillouin Zone that are protected under lattice distortions. It either measures the circulation of isolated Weyl point nodes $C$ encloses~\cite{po2016phonon, rocklin2016mechanical} or characterizes the topological polarization if $C$ is a non-contractible loop across the torus~\cite{kane2014topological}. But remarkably, {\it for a generic model of spin origami} we find $\text{Det}[ {\bf \mathcal{R}}({\bf k})]$ is a real number up to an overall constant phase in the Brillouin zone (BZ). It obeys the mysterious ``realness'' condition previously observed for the rigidity matrices of triangulated mechanical origami~\cite{chen2016topological}. The winding numbers $w(C)$ therefore vanish for all $C$. After eliminating a constant phase by choosing a gauge, however, this realness condition defines another topological number:
\begin{equation}
  \eta({\bf k}) = \text{sign}\ \text{Det} [\mathcal{R}({\bf k})].
  \label{eq:topo}
\end{equation}
It demands two regions in the BZ with different $\eta({\bf k})$ are separated by a line of zero modes -- the topological Weyl line nodes. We illustrate this in the supplemental materials by generating periodic origami and observing how these line nodes move and can vanish pairwise. So just by computing $\eta({\bf k})$ we can learn a lot about the zero modes: while they may be lifted by distortions [see Fig.~\ref{figg4} (c)], a generic nonflattenable origami typically still has topological Weyl line nodes in its spin wave dispersion. The Dirac line nodes must then somehow be pairs of these Weyl line nodes.

To explain the double degeneracy, we have carried out a symmetry analysis in the Supplementary Material. We now know adding a symmetry can eliminate topology and create new topology. Specifically for ${\rm Cs}_2{\rm CeCu}_3{\rm F}_{12}$, whose triangular faces pair up to create diamond shapes, its point group symmetry explains the numerically observed double degeneracy by playing a role analogous to Kramers degeneracy in a metal. By plotting the 12 spins within the unit cell with tails at a common origin we have uncovered precisely such a symmetry. We find the point group has both unitary and antiunitary symmetries which guarantee that we can place the rigidity matrix in a block diagonal form with two $12\times12$ blocks each with just real numbers as their elements. The determinant then becomes $\text{Det}[{\mathcal R}({\bf k})] = \text{Det}[{\mathcal R}_+({\bf k})]\text{Det}[{\mathcal R}_-({\bf k})]$ where not only $\text{Det}[{\mathcal R}({\bf k})]$ is real, but also $\text{Det}[{\mathcal R}_\pm({\bf k})]$. We can then define new topological invariants $\eta_\pm({\bf k}) = \text{sign}\ \text{Det}[{\mathcal R}_\pm({\bf k})]$ with $\eta({\bf k}) = \eta_+({\bf k})\eta_-({\bf k})$. A plot of $\eta_+({\bf k})$ is shown in Fig.~\ref{figg4} evincing the effects of distortion that splits the Dirac line nodes into Weyl type. The point group symmetry further demands they both change sign if one of them changes sign so that $\eta({\bf k})$ never changes sign (a loss of topology) and any line nodes are doubly degenerate (a new topology). Hence, by identifying the full point group symmetry and its antiunitary character, we have explained the topological protection of the double degenerate line nodes. 

In summary we have identified broad classes of KHAF, including two experimentally available fluoride compounds, whose degenerate ground states can be mapped onto the folding motions of origami sheets. The geometry, symmetry and topology of these mechanical analogs explicates how seemingly comparable spin interactions can either preserve or destroy the extensive frustration, or even give rise to novel Dirac line nodes. This mapping extends the original spin origami concept to permit new notions of folding and straining structured mechanical sheets. New results in topological mechanical metamaterials suggest that other magnetic systems may yet realize exotic gapless modes on the boundary and Weyl point nodes in the bulk.

\noindent
{\underline{\sl{Acknowledgments :-}}} We thank Tom C. Lubensky, James P. Sethna, and Itai Cohen for useful discussion. KR and MJL acknowledge supported in part by the National Science Foundation under Grant No. NSF PHY17-48958. DZR gratefully acknowledge support from the the ICAM postdoctoral fellowship, the Bethe/KIC Fellowship, and the National Science Foundation Grant No. NSF DMR-1308089..


\bibliography{reference_ordered}
\end{document}


\title{Topology and geometry of spin origami: supplementary material}

\author{Krishanu Roychowdhury}
\affiliation{Laboratory of Atomic And Solid State Physics, Cornell University, Ithaca, NY 14853.}
\affiliation{Kavli Institute for Theoretical Physics, University of California, Santa Barbara, CA 93106-4030.}
\author{D. Zeb Rocklin}
\affiliation{Laboratory of Atomic And Solid State Physics, Cornell University, Ithaca, NY 14853.}
\affiliation{School of Physics, Georgia Institute of Technology, Atlanta, GA 30332.}
\author{Michael J. Lawler\footnote{Corresponding author.}}
\affiliation{Laboratory of Atomic And Solid State Physics, Cornell University, Ithaca, NY 14853.}
\affiliation{Kavli Institute for Theoretical Physics, University of California, Santa Barbara, CA 93106-4030.}
\affiliation{Department of Physics, Binghamton University, Binghamton, NY, 13902.}

\maketitle

\section{Derivation of the star condition}\label{secone} 

\begin{figure}
\centering
 \includegraphics[width=10cm]{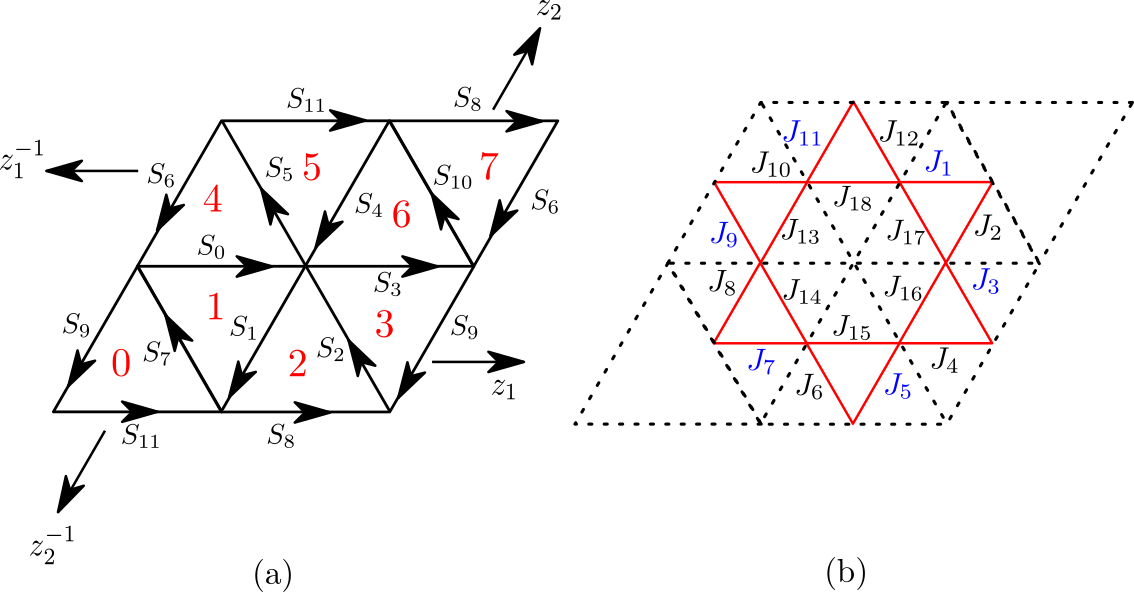}
 \caption{(a) The spin assignment with arrows denoting the orientation associated with the definition of the spin vector -- positive along the direction of the arrow. (b) The corresponding kagome Star of David with exchange interactions marked is inscribed inside the origami unit cell.}
\label{fig_latt1}
\end{figure}

The following table is provided to make connections between the set of spins and the set of triangles in the origami sheet [Supplementary Fig.~\ref{fig_latt1} (a)] with the exchange interactions marked on the corresponding kagome Star of David [Supplementary Fig.~\ref{fig_latt1} (b)].
\begin{center}
 \begin{tabular}{|c|c|c|} 
 \hline
 Triangle label & Spins in the triangle & exchange interactions in kagome \\  
 \hline
 1 & $\spin_0$, $\spin_1$, $\spin_7$ & $J_7$, $J_{8}$, $J_{14}$ \\
 \hline
 2 & $\spin_1$, $\spin_2$, $\spin_8$ & $J_5$, $J_{6}$, $J_{15}$  \\
 \hline
 3 & $\spin_2$, $\spin_3$, $\spin_9$ & $J_3$, $J_{4}$, $J_{16}$  \\
 \hline
 4 & $\spin_0$, $\spin_5$, $\spin_{6}$ & $J_9$, $J_{10}$, $J_{13}$  \\
 \hline
 5 & $\spin_4$, $\spin_5$, $\spin_{11}$ & $J_{11}$, $J_{12}$, $J_{18}$  \\  
 \hline
 6 & $\spin_3$, $\spin_4$, $\spin_{10}$ & $J_{1}$, $J_{2}$, $J_{17}$  \\  
 \hline
\end{tabular}
\end{center}
Now consider the two adjacent triangles with label 4 and 1 in Supplementary Fig.~\ref{fig_latt1} (a) which are associated with the constraints:
\begin{equation}
\spin_{\triangle_4} = \ell^{\triangle_4}_0 \spin_0 + \ell^{\triangle_4}_5 \spin_5 + \ell^{\triangle_4}_6 \spin_6 = 0~~;~~\spin_{\triangle_1} = \ell^{\triangle_1}_0 \spin_0 + \ell^{\triangle_1}_1 \spin_1 + \ell^{\triangle_1}_{7} \spin_{7} = 0. 
\end{equation}
Following $J_{ij}=J_{\triangle_m} \ell^{\triangle_m}_i\ell^{\triangle_m}_j$ from Eq.~\ref{txt-eq:ham1} of the main text, we find that for the common spin $\spin_0$,
\begin{equation}
 \ell^{\triangle_4}_0=\sqrt{J_{9}J_{13}/J_{\triangle_4} J_{10}}~~;~~\ell^{\triangle_1}_0=\sqrt{J_{8}J_{14}/J_{\triangle_1} J_7}.
\end{equation}
Construction of an origami sheet follows from equating these two lengths which we refer to as a {\it length consistency condition}. This leads to 
\begin{equation}
 J_{\triangle_1} = \bigg(\frac{J_8J_{10}J_{14}}{J_7J_9J_{13}}\bigg)J_{\triangle_4}.
 \label{eq:cond1}
\end{equation}
Continuing this way, for triangle 1 and 2 we could apply the length consistency condition to the edge that corresponds to the common spin $\spin_1$ obtaining
\begin{equation}
 J_{\triangle_2} = \bigg(\frac{J_6J_{8}J_{15}}{J_5J_7J_{14}}\bigg)J_{\triangle_1}.
 \label{eq:cond2}
\end{equation}
Considering the consecutive adjacent pairs of triangles in a similar way, we get
\begin{equation}
 \begin{split}
  J_{\triangle_3} = \bigg(\frac{J_4J_{6}J_{16}}{J_3J_5J_{15}}\bigg)J_{\triangle_2}, \\
  J_{\triangle_6} = \bigg(\frac{J_2J_{4}J_{17}}{J_1J_3J_{16}}\bigg)J_{\triangle_3}, \\
  J_{\triangle_5} = \bigg(\frac{J_2J_{12}J_{18}}{J_1J_{11}J_{17}}\bigg)J_{\triangle_6}.
 \end{split}
 \label{eq:cond3}
\end{equation}
In other words, we get a ratio for the adjacent triangle 4 and 5 by demanding length consistency for all the edges representing $\spin_j$ for $j=0$ to $4$
\begin{equation}
 J_{\triangle_5} = \bigg(\frac{J_2J_{12}J_{18}}{J_1J_{11}J_{17}}\bigg)\bigg(\frac{J_2J_{4}J_{17}}{J_1J_3J_{16}}\bigg)\bigg(\frac{J_4J_{6}J_{16}}{J_3J_5J_{15}}\bigg)\bigg(\frac{J_6J_{8}J_{15}}{J_5J_7J_{14}}\bigg)\bigg(\frac{J_8J_{10}J_{14}}{J_7J_9J_{13}}\bigg) J_{\triangle_4}.
 \label{eq:cond4}
\end{equation}
These two triangles share the common spin $\spin_5$, and applying length consistency for that edge implies
\begin{equation}
 J_{\triangle_5} = \bigg(\frac{J_9J_{11}J_{18}}{J_{10}J_{12}J_{13}}\bigg)J_{\triangle_4}.
 \label{eq:cond5}
\end{equation}
Comparing Eq.~\ref{eq:cond5} with Eq.~\ref{eq:cond4} we get
\begin{equation}
  (J_{1}J_{3}J_{5}J_{7}J_{9}J_{11})^2=(J_{2}J_{4}J_{6}J_{8}J_{10}J_{12})^2,
 \label{eq:cond6}
\end{equation}
from which the star condition (presented in Eq.~\ref{txt-eq:star} of the main text) follows as all $J$'s are positive. The condition ensures closure of the origami hexagon comprising six adjacent triangles joined at a common vertex. As such, when it is satisfied for each Star of David of the spin system, the ground state configurations may be mapped onto the ground state configurations of a particular nonflattenable triangulated origami sheet. \\


\section{The rigidity matrix and spin wave dispersions in ${\rm Cs}_2{\rm CeCu}_3{\rm F}_{12}$}\label{sectwo}

The quantities $J_{\triangle_m}$ mentioned above are merely scale factors that ensures the length consistency on each edge of the origami sheet. They do not alter the zero modes but modify the finite frequencies of the spin waves; in that sense, $\tilde{{\bf S}}_{\triangle_m}\equiv\sqrt{J_{\triangle_m}}{\bf S}_{\triangle_m}$ is equally good a constraint function to consider. The spin Hamiltonian in this language is 
\begin{equation}
 H = \frac{1}{2}\sum_{\triangle_m} \sum_{j\in\triangle_m} (a^{\triangle_m}_j{\bf S}_j)^2~~;~~ a^{\triangle_m}_j=\sqrt{J_{ij}J_{ik}/J_{jk}}~~ \text{for triangle}~~\triangle_m =\langle ijk\rangle. 
\end{equation}
Below we list down the constraints using which the rigidity matrix for the KHAF in ${\rm Cs}_2{\rm CeCu}_3{\rm F}_{12}$ is calculated [see Supplementary Fig.~\ref{fig_latt2} (a) and (b)].
\begin{equation}
 \begin{split}
  a^{\triangle_0}_7{\bf S}_7 + a^{\triangle_0}_9{\bf S}_9 + a^{\triangle_0}_{11}{\bf S}_{11} &= 0~~;~~ a^{\triangle_0}_7=\sqrt{J_2}, a^{\triangle_0}_9=\sqrt{J_2}, a^{\triangle_0}_{11}=\sqrt{J_3^2/J_2} \\
  a^{\triangle_1}_7{\bf S}_7 + a^{\triangle_1}_0{\bf S}_0 + a^{\triangle_1}_{1}{\bf S}_{1} &= 0~~;~~ a^{\triangle_1}_7=\sqrt{J_1}, a^{\triangle_1}_0=\sqrt{J_4^2/J_1}, a^{\triangle_1}_{1}=\sqrt{J_1} \\
  a^{\triangle_2}_2{\bf S}_2 + a^{\triangle_2}_1{\bf S}_1 + a^{\triangle_2}_{8}{\bf S}_{8} &= 0~~;~~ a^{\triangle_2}_2=\sqrt{J_2}, a^{\triangle_2}_1=\sqrt{J_2}, a^{\triangle_2}_{8}=\sqrt{J_3^2/J_2} \\
  a^{\triangle_3}_2{\bf S}_2 + a^{\triangle_3}_9{\bf S}_9 + a^{\triangle_3}_{3}{\bf S}_{3} &= 0~~;~~ a^{\triangle_3}_2=\sqrt{J_1}, a^{\triangle_3}_9=\sqrt{J_1}, a^{\triangle_3}_{3}=\sqrt{J_4^2/J_1} \\
  a^{\triangle_4}_0{\bf S}_0 + a^{\triangle_4}_6{\bf S}_6 + a^{\triangle_4}_{5}{\bf S}_{5} &= 0~~;~~ a^{\triangle_4}_0=\sqrt{J_4^2/J_1}, a^{\triangle_4}_6=\sqrt{J_1}, a^{\triangle_4}_{5}=\sqrt{J_1} \\
  a^{\triangle_5}_4{\bf S}_4 + a^{\triangle_5}_5{\bf S}_5 + a^{\triangle_5}_{11}{\bf S}_{11} &= 0~~;~~ a^{\triangle_5}_4=\sqrt{J_2}, a^{\triangle_5}_5=\sqrt{J_2}, a^{\triangle_5}_{11}=\sqrt{J_3^2/J_2} \\
  a^{\triangle_6}_3{\bf S}_3 + a^{\triangle_6}_4{\bf S}_4 + a^{\triangle_6}_{10}{\bf S}_{10} &= 0~~;~~ a^{\triangle_6}_3=\sqrt{J_4^2/J_1}, a^{\triangle_6}_4=\sqrt{J_1}, a^{\triangle_6}_{10}=\sqrt{J_1} \\
  a^{\triangle_7}_6{\bf S}_6 + a^{\triangle_7}_8{\bf S}_8 + a^{\triangle_7}_{10}{\bf S}_{10} &= 0~~;~~ a^{\triangle_7}_6=\sqrt{J_2}, a^{\triangle_7}_8=\sqrt{J_3^2/J_2}, a^{\triangle_7}_{10}=\sqrt{J_2}. \\
 \end{split}
\end{equation}
Let us fix one of the scale factors, say $J_{\triangle_1}=J_{\triangle}$. All other scale factors can be derived from $J_{\triangle_1}$ by demanding length consistency on the edges of the origami which yields
\begin{equation}
 J_{\triangle_0}=J_{\triangle}J_2/J_1, J_{\triangle_2}=J_{\triangle}J_2/J_1, J_{\triangle_3}=J_{\triangle}, J_{\triangle_4}=J_{\triangle}, J_{\triangle_5}=J_{\triangle}J_2/J_1, J_{\triangle_6}=J_{\triangle}, J_{\triangle_7}=J_{\triangle}J_2/J_1. 
\end{equation}
The nonflattenable origami sheet to which the ground state of KHAF in ${\rm Cs}_2{\rm CeCu}_3{\rm F}_{12}$ corresponds has three distinct edge lengths $l_a$, $l_b$ and $l_c$ [see Supplementary Fig.~\ref{fig_latt2} (c)] that are related to the interactions $J_1$, $J_2$, $J_3$ and $J_4$ as
\begin{equation}
 l_a=\frac{J_3}{J_2}\sqrt{\frac{J_1}{J_\triangle}}~~;~~l_b=\sqrt{\frac{J_1}{J_\triangle}}~~;~~l_c=\frac{J_4}{J_1}\sqrt{\frac{J_1}{J_\triangle}}, 
\end{equation}
such that $l_a/l_b=J_3/J_2$ and $l_c/l_b=J_4/J_1$. In our calculations, WLOG we set $J_\triangle$ to be the largest energy scale of the problem which is $J_1$. This sets $l_b=1$, $l_a=J_3/J_2$ and $l_c=J_4/J_1$ in the origami (the values of $J_{1,2,3,4}$ are provided in the main text).

\begin{figure}
\centering
 \includegraphics[width=16cm]{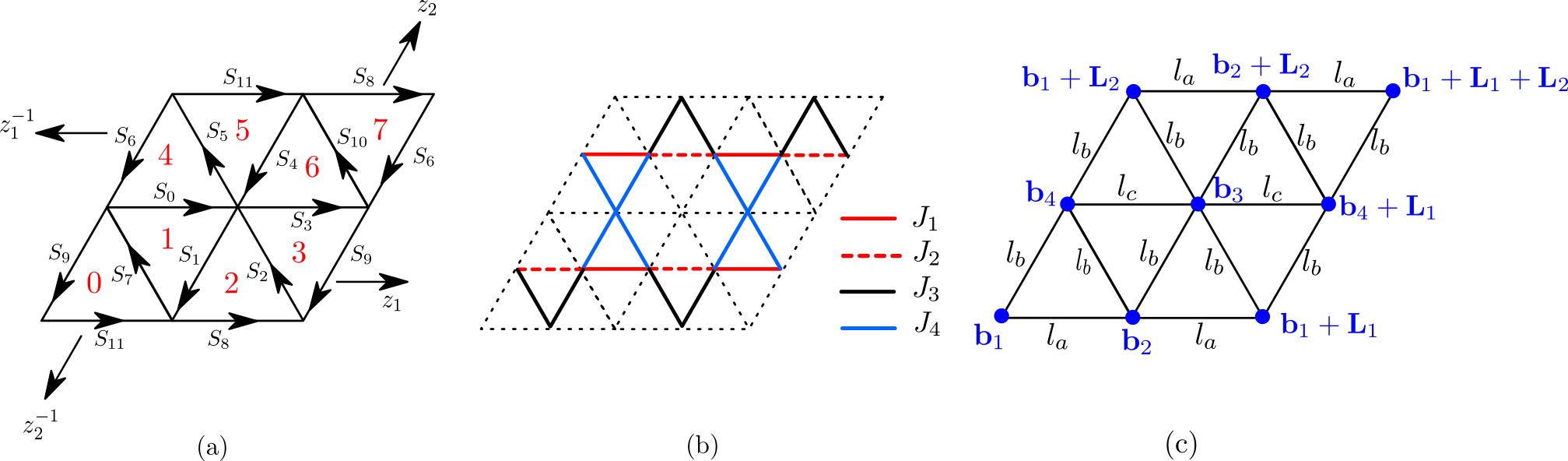}
 \caption{(a) The origami unit cell for ${\rm Cs}_2{\rm CeCu}_3{\rm F}_{12}$ with the spins and faces labelled. (b) The corresponding kagome Star of David for ${\rm Cs}_2{\rm CeCu}_3{\rm F}_{12}$ with exchange interactions marked is inscribed inside the origami unit cell. (c) The origami unit cell with edge lengths and vertices specified.}
\label{fig_latt2}
\end{figure}

The origami unit cell contains four distinct vertices characterized by four basis vectors ${\bf b}_1$, ${\bf b}_2$, ${\bf b}_3$ and ${\bf b}_4$ and two lattice vectors ${\bf L}_1$ and ${\bf L}_2$ [see Supplementary Fig.~\ref{fig_latt2} (c)]. The unit vectors along the edges are the spins given in Supplementary Fig.~\ref{fig_latt2} (a) whose orientations are obtained through the following procedure. The interaction pattern in ${\rm Cs}_2{\rm CeCu}_3{\rm F}_{12}$ [Supplementary Fig.~\ref{fig_latt2} (b)] demands, for a given basis, the edge from $\bvec_1$ to $\bvec_4$ and the one from $\bvec_2$ to $\bvec_3$ is equal. The vertices at ${\bf b}_1$, ${\bf b}_2$, and ${\bf b}_4$ form an isosceles triangle [triangle 0 in Supplementary Fig.~\ref{fig_latt2} (a)] which can be defined entirely on a plane with ${\bf b}_1=(0,0,0)$, ${\bf b}_2=(l_a,0,0)$ and ${\bf b}_4=(l_a/2,\sqrt{l_b^2-l_a^2/4},0)$. Now the origami configuration demands the following set of equations
\begin{equation}
\begin{split}
 || {\bf b}_3 - {\bf b}_4 || &= l_c \\ 
 || {\bf b}_3 - {\bf b}_2 || &= l_b \\ 
 ||{\bf b}_1 + {\bf L}_1 - {\bf b}_2 || &= l_a \\ 
 ||{\bf b}_1 + {\bf L}_1 - {\bf b}_3 || &= l_b \\ 
 ||{\bf b}_4 + {\bf L}_1 - {\bf b}_3 || &= l_c \\ 
\end{split}
\end{equation}
to solve for ${\bf b}_3$ and ${\bf L}_1$, which together have six variables ($x$, $y$ and $z$ component of each). This leads us to define a tuning parameter ${\bf b}_3^z=b_0$ (in Fig.~\ref{txt-figg3} of the main text) in a range $[-0.12,0.12]$ to allow for real solutions, thus, defining a one-dimensional family of configurations for the nonflattenable sheet. We further notice that setting ${\bf L}_2={\bf b}_3+{\bf b}_4-{\bf b}_1-{\bf b}_2$ satisfies all other constraints on the sheet and lead to the symmetries associated with the spin configuration resulting in diamond-shaped faces of the origami noted in the main text. For a given $b_0$, the spin configuration of the ground state is then uniquely specified through the vectors ${\bf b}_1$, ${\bf b}_2$, ${\bf b}_3$, ${\bf b}_4$, ${\bf L}_1$, and ${\bf L}_2$. Distortions can be made to the sheet by introducing a parameter $t_0$ to modify the definition of ${\bf L}_2$ as ${\bf L}_2={\bf b}_3+{\bf b}_4-{\bf b}_1-t_0{\bf b}_2$. When $t_0$ is set away from 1, it produces a combined effects of shearing and stretching the sheet as noted in Fig.~\ref{txt-figg4} of the main text. \\

The directions of the 12 spins in the unit cell of the origami are the following [see Supplementary Fig.~\ref{fig_latt2} (a) and (c)]
\begin{equation}
\begin{split}
 {\bf S}_0    &= ({\bf b}_3 - {\bf b}_4)/|| {\bf b}_3 - {\bf b}_4 || \\ 
 {\bf S}_1    &= ({\bf b}_2 - {\bf b}_3)/|| {\bf b}_2 - {\bf b}_3 || \\
 {\bf S}_2    &= ({\bf b}_3 - {\bf b}_1 - {\bf L}_1)/|| {\bf b}_3 - {\bf b}_1 - {\bf L}_1 || \\
 {\bf S}_3    &= ({\bf b}_4 + {\bf L}_1 - {\bf b}_3)/|| {\bf b}_4 + {\bf L}_1 - {\bf b}_3 || \\
 {\bf S}_4    &= ({\bf b}_3 - {\bf b}_2 - {\bf L}_2)/|| {\bf b}_3 - {\bf b}_2 - {\bf L}_2 || \\
 {\bf S}_5    &= ({\bf b}_1 + {\bf L}_2 - {\bf b}_3)/|| {\bf b}_1 + {\bf L}_2 - {\bf b}_3 || \\
 {\bf S}_6    &= ({\bf b}_4 - {\bf b}_1 - {\bf L}_2)/|| {\bf b}_4 - {\bf b}_1 - {\bf L}_2 || \\
 {\bf S}_7    &= ({\bf b}_4 - {\bf b}_2)/|| {\bf b}_4 - {\bf b}_2 || \\
 {\bf S}_8    &= ({\bf b}_1 + {\bf L}_1 - {\bf b}_2)/|| {\bf b}_1 + {\bf L}_1 - {\bf b}_2 || \\
 {\bf S}_9    &= ({\bf b}_1 - {\bf b}_4)/|| {\bf b}_1 - {\bf b}_4 || \\
 {\bf S}_{10} &= ({\bf b}_2 - {\bf b}_4 + {\bf L}_2 - {\bf L}_1)/|| {\bf b}_2 - {\bf b}_4 + {\bf L}_2 - {\bf L}_1 || \\
 {\bf S}_{11} &= ({\bf b}_2 - {\bf b}_1)/|| {\bf b}_2 - {\bf b}_1 || \\ 
\end{split}
\end{equation}
As these spins are unit vectors, each of them can be parametrized by two canonical variables, $x^j$ and $p_j$ as
\begin{equation}
  {\bf{S}}_j=\big\{ \cos(q^j)\sqrt{1-p^2_j},~\sin(q^j)\sqrt{1-p^2_j},~p_j \big\}, \nonumber
\end{equation}
with the Poisson bracket $\{q^i,p_j\}=\delta^i_j$. The rigidity matrix ${\mathcal R}$ is defined by linearizing the constraints about the ground state as $S_{\triangle\alpha} = {\mathcal R}_{\triangle\alpha,j\mu}x^{j\mu}$ (see Eq.~\ref{txt-eq:rig} of the main text), where $j\in\triangle$ and $x^{j\mu}\in(q^j,p_j)$. \\

An explicit construction of ${\mathcal R}$ follows by considering the basis 
\begin{equation}
 \tau_1= [q_0,q_1,\cdots,\\ q_{11},p_0,p_1,\cdots,p_{11}]^T \nonumber \\
\end{equation}
corresponding to the twelve spins $\spin_0,\spin_1,\cdots,\spin_{11}$ in the unit cell and 
\begin{equation}
 \tau_2=[\triangle_0^x,\cdots,\triangle_7^x,\triangle_0^y,\cdots,\triangle_7^y,\triangle_0^z,\cdots,\triangle_7^z]^T \nonumber \\
\end{equation}
corresponding to the eight faces in the unit cell shown in Supplementary Fig.~\ref{fig_latt2} (a) such that $\tau_2=\mathcal{R}\cdot \tau_1$. \\

To translate to the momentum space, one needs to consider a Fourier transformation of $\mathcal{R}$. We chose the convention
\begin{equation}
  {\mathcal R}_{\triangle\alpha,j\mu} = \frac{1}{\sqrt{N}} \sum_{\bf{k}} {\mathcal R}_{{\bf{D}}\alpha,{\bf{d}}\mu} ({\bf{k}}) ~ e^{\iota {\bf{k}}\cdot({\bf{R}}_i-{\bf{R}}'_i)},
  \label{fourier_A}
\end{equation}
where $N$ is the number of sites in a unit cell; ${\bf{R}}_i$ and ${\bf{R}}'_i$ denote the position vectors of the unit cells which contain the triangle $\triangle$ at ${\bf{R}}_i+{\bf{D}}$ and the site $j$ at ${\bf{R}}'_i+{\bf{d}}$ respectively. Our particular choice of unit cells is shown in Fig. \ref{fig_latt2} (a). The sites associated with phases $z_1=e^{\iota k_1a}$, $z_2=e^{\iota k_2a}$, $z_1^{-1}$, $z_2^{-1}$ lie in neighboring unit cells. Here $k_1a = {\bf k}\cdot{\bf T}_1$, $k_2a = {\bf k}\cdot{\bf T}_2$, ${\bf T}_1$ and ${\bf T}_2$ are Bravais lattice vectors for the ordering pattern. With this choice of Fourier transform convention and unit cells, we find that ${\rm Det}[\mathcal{R}({\bf k})]$ is real everywhere in the Brillouin zone.\\

Dispersions of the spin waves (the magnon frequencies) are associated with the equation of motion of $x^{i\mu}$. Considering $J^{\triangle\alpha,\triangle'\beta}=\delta^{\triangle\alpha,\triangle'\beta}$ for models with only nn interactions these equations read
\begin{equation}
 \dot{x}^{i\mu} = \sigma^{i\mu,j\nu} {\mathcal R}^{T}_{j\nu,\triangle\alpha}
\delta^{\triangle\alpha,\triangle'\beta}{\mathcal R}_{\triangle'\beta,k\lambda} x^{k\lambda}\to \dot{\bf x} = {\bf \sigma}{\bf \mathcal R}^T{\bf \mathcal R}{\bf x},
 \label{eigen1}
\end{equation}
where $\sigma^{i\mu,j\nu}=\{x^{i\mu},x^{j\nu}\}=\delta^{ij}\epsilon^{\mu\nu}$ is the Poisson bracket tensor with $\epsilon^{\mu\nu}$ the two-dimensional Levi-Civita tensor. Using the momentum space representation following Eq.~\ref{fourier_A}, the magnon frequencies $\omega({\bf{k}})$ are calculated [and plotted in Fig.~\ref{txt-figg4} (d) of the main text] by diagonalizing the matrix ${\bf{\sigma}}{\mathcal R}^T(-{\bf{k}}) {\mathcal R}({\bf{k}})$, cf. Eq.~\ref{eigen1}.\\


\section{Influence of Gaussian curvature of the origami sheet on magnetic structure}\label{secthree}

\begin{figure}
\centering
 \includegraphics[width=8cm]{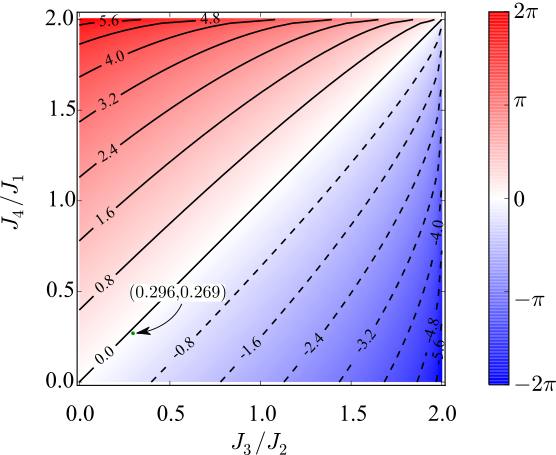}
 \caption{The plot of $\mathcal{G}$ in the space of interaction ratios $J_3/J_2$ and $J_4/J_1$ (see Eq.~\ref{eq:nonflat}) with contours of specific $\mathcal{G}$ values [the solid(dashed) contours represent positive(negative) values] shown. The green point represents the ${\rm Cs}_2{\rm CeCu}_3{\rm F}_{12}$ compound which lies close to the $\mathcal{G}=0$ line. Tuning the interactions in ${\rm Cs}_2{\rm CeCu}_3{\rm F}_{12}$ would drive a topological phase transition (change in the topology of the origami sheet) characterized by the change of the sign of $\mathcal{G}$.}
\label{fig_latt3}
\end{figure}

Any smooth curved surface embedded in three-dimensional space has Gaussian curvature, the product of the two principle curvatures (inverse radii of curvature) at that point~\cite{kamien2002geometry}. The origami surfaces considered in the present work are not smooth, but exist as the limit of smooth surfaces with tubular curves along edges and spherical caps at vertices. These caps are then the only sources of Gaussian curvature on the sheet, and while their curvature diverges, the integral of curvature over the cap has a well-defined limit equal to the \emph{angle deficit} 

\begin{align}
\label{eq:gauss}
\mathcal{G}_{\hexagon} = 2\pi - \sum_{\langle ij \rangle \in \hexagon} \theta_{ij},
\end{align}

\noindent where the sum is over interior angles of the triangles adjoining the vertex~\cite{calladine1988theory}. Only when this curvature vanishes can all the edges adjoining the vertex become coplanar, a necessary condition for a zero mode to become localized at the vertex. This mode consists of the vertex displacing transverse to the plane, or to the six spins rotating transverse to their own plane in the magnetic system. For the nonflattenable origami sheet corresponding to the ground state of the KHAF model in ${\rm Cs}_2{\rm CeCu}_3{\rm F}_{12}$, vertices have a finite Gaussian curvature $\pm \mathcal{G}$ with

\begin{equation}
 \mathcal{G} = 4\cos^{-1}\frac{J_3}{2J_2} - 4\cos^{-1}\frac{J_4}{2J_1},
 \label{eq:nonflat}
\end{equation}

\noindent which arises because of the two distinct types of kagome triangles, consequently two distinct types of origami triangles whose shapes are determined by the ratios $J_4/J_1$ (the kagome triangle consisting of two $J_4$ and one $J_1$) and $J_3/J_2$ (the kagome triangle consisting of two $J_3$ and one $J_2$) as shown in Supplementary Fig.~\ref{fig_latt2} (b). When the ratios are the same, it leads to a flat sheet (locally flat at each vertex) with $\mathcal{G}=0$. 

One can make distortions to the compound to track the variation of $\mathcal{G}$ as the interactions change (there exist a number of experiments on kagome systems where tuning the exchange interactions have been made possible by applying uniaxial stress or pressure, see the main text for references). A plot of $\mathcal{G}$ in the space of $J_3/J_2$ and $J_4/J_1$ reveals interesting features as shown in Supplementary Fig.~\ref{fig_latt3}. The plot suggests that tuning those parameters appropriately could drive a transition in which the quantity $\mu\equiv{\rm sgn}(\mathcal{G})$ flips. Consequently, the origami sheet passes from one nonflattenable state to another through a locally flat sheet (locally flat at each vertex). Emergence of zero modes corresponding to that flat sheet at the transition point (when the ratios $J_3/J_2$ and $J_4/J_1$ are the same) indicates a topological phase transition characterized by $\mu$ signifying it a topological invariant for the KHAF model. 

While the curvatures of individual vertices are local, geometric properties, the curvature of the whole unit cell is fixed by the fundamental topology of the two-dimensional surface. The celebrated Gauss-Bonnet theorem~\cite{kamien2002geometry} holds that for any compact two-dimensional manifold $M$, its Euler characteristic $\chi$ can be related to the integrals of its Gaussian curvature $K$ over its bulk and its geodesic curvature $k_g$ around its boundary:

\begin{align}
2\pi \chi = \iint K d A + \oint k_g d \ell.
\end{align}

\noindent We choose as our boundary a counter-clockwise path around one or more unit cells of the origami sheet. This disk-type patch has Euler characteristic one and because the boundary is along the periodic cell in opposite directions, most of its curvature terms cancel out. What remains are the rotations that account for the full revolution of the orientation vector by $2\pi$, canceling out with the contribution from the nonzero Euler characteristic. As a result, we have that the Gaussian curvature integrated over the unit cell must vanish:

\begin{align}
\iint_\textrm{cell} K d A = \sum_{\hexagon\in\textrm{cell}} \mathcal{G}_{\hexagon} = 0.
\end{align}

\noindent Thus, topology restricts the geometry of the origami such that the total angle deficit around the unit cell must vanish. This is particularly striking for ${\rm Cs}_2{\rm CeCu}_3{\rm F}_{12}$, which has only two types of vertex, which then must necessarily have opposite angle deficits.

\section{Topological Weyl line nodes in periodic spin origami}\label{secfour}
We generate generic periodic origami, origami with no special point group symmetry just a translational symmetry and study their Weyl line nodes. We do so by solving for origami sheets of a corresponding kagome antiferromagnet with no special conditions on their exchange interactions other than the star condition discussed in the main text. As shown in Supplementary Fig.~\ref{fig_weyl}, these line nodes move around as we tune parameters and can even annihilate and are characterized by the topological invariant $\eta({\bf k})$ as expected from the theory presented in the main text. 

\begin{figure*}
\centering
 \includegraphics[width=17.6cm]{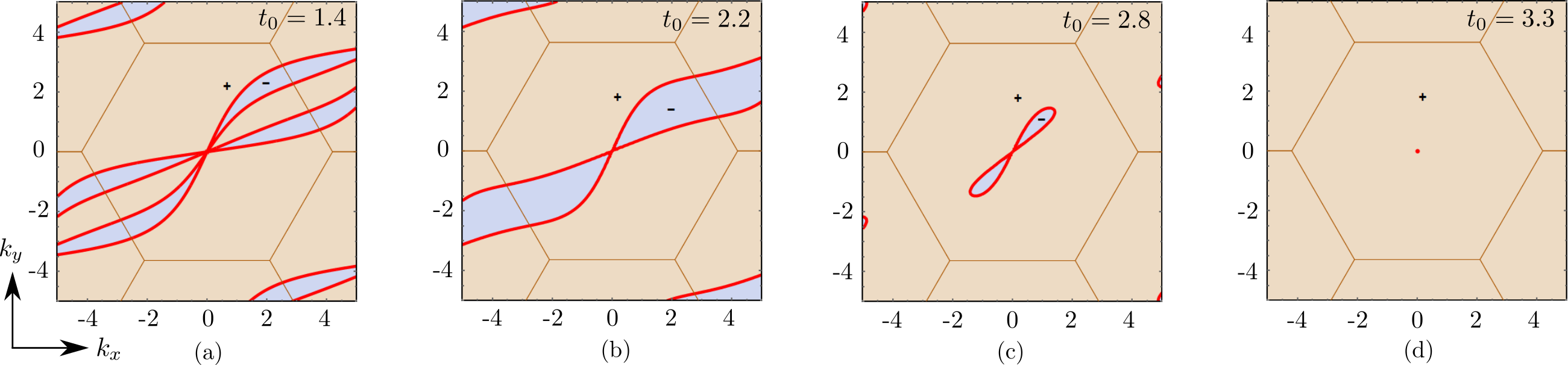}
 \caption{(a) Weyl line nodes (thick red lines) appearing in the plot of the topological number $\eta$ (Eq.~\ref{txt-eq:topo} of the main text) over the Brillouin zone of a generic KHAF corresponding to a nonflattenable origami surface. The lines separate the two regions of positive (yellow) and negative (blue) values of $\eta$ denoted by `+' and `-' respectively. The quantity $t_0$ produces distortions involving the coupling constants of the spin model in Eq.~\ref{txt-eq:ham1} of the main text. A continuous deformation of the sheet by tuning a parameter (we call $t_0$) sequentially leads to (a)$\rightarrow$(b)$\rightarrow$(c)$\rightarrow$(d) in which first a pair of line nodes vanishes in (b) (at the Brillouin zone boundary the vanishing takes place at the $M$ points), then the `-' region gradually shrinks down as in (c) till in (d), the nodes disappear altogether except the $\Gamma$ point which remains always gapless owing to the global spin rotation symmetry.}
 \label{fig_weyl} 
\end{figure*}


\section{Symmetry analysis of ${\rm Cs}_2{\rm CeCu}_3{\rm F}_{12}$ ground state}\label{secfive}

The unexpected Dirac line nodes (doubly degenerate line nodes) appear together with a point group symmetry in our analysis of ${\rm Cs}_2{\rm CeCu}_3{\rm F}_{12}$. The generic nonflattenable origami presented in Fig.~\ref{txt-figg2} (d) of the main text has singly degenerate line nodes and no point group symmetry. So it is natural to expect the double degeneracy follows from the added symmetry. Here we prove that this is the case.\\

To reveal the symmetry of ${\rm Cs}_2{\rm CeCu}_3{\rm F}_{12}$'s periodic nonflattenable origami sheet, we have plotted in Supplementary Fig.~\ref{figM1}, one component of its spin structure in real space as well constructed a ``sphere plot" of the 12 spins in its unit cell. The sphere plot places the tail of each spin at the origin of three dimensional Euclidean space and illuminates the possibility of a symmetry in spin space that might be combined with a space group transformation.\\

\begin{figure}
\centering
 \includegraphics[width=6cm]{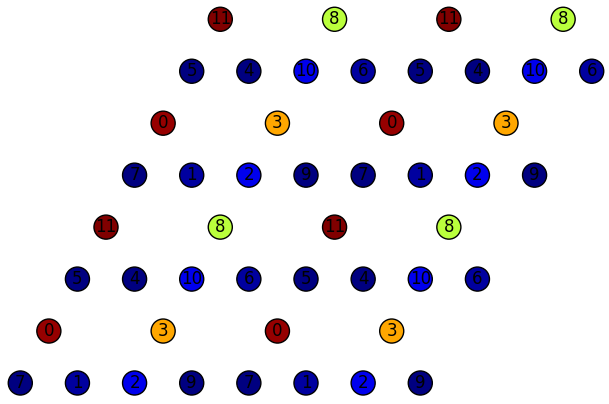}
 \includegraphics[width=6cm]{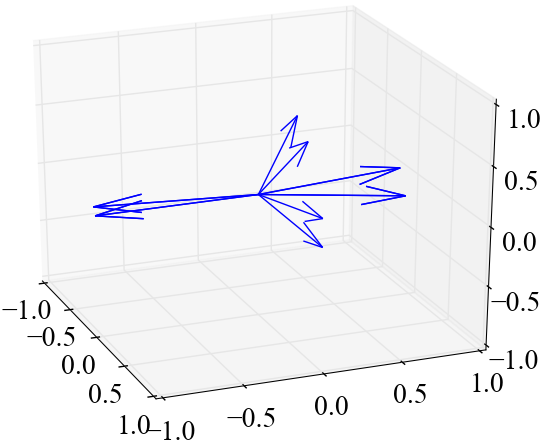}
 \caption{Plotting spin ground state pattern to reveal symmetries. {\bf Left}: the $x$-component of the spin vectors plotted on the kagome lattice. This reveals the $C_2^{site}$ symmetry about sites 0, 3, 8 and 11. {\bf Right}: a ``sphere plot'' of the spins within the unit cell consisting of placing the tail of each spin at the origin of a 3 dimensional Euclidean space. This reveals a mirror symmetry $M$ in spin space.}
\label{figM1}
\end{figure}

The plots show the following symmetries exist:

\begin{itemize}
\setlength{\itemindent}{0.63cm}
\item A 180$^o$ $C_2^{site}$ spatial rotation symmetry about four of the sites within the unit cell that sends ${\bf r} \equiv {\bf R} + {\bf d} \to -{\bf r}$. These are all the same transformation up to translations (i.e. the same transformation within the point group).
\item A mirror symmetry $M$ in spin space followed by either a second $C_2^{hex}$ rotation about the center of a hexagon or a translation $T_1^{1/2}$ by half the unit cell of the ordering pattern.
\end{itemize}

In total, the three transformations $C_2^{site}$, $MC_{2}^{hex}$ and $MT_1^{1/2}$ together with the identity make the group $Z_2\times Z_2$. Because all variables are real in real space, there is also a complex-conjugation symmetry $K$ as pointed out in Ref.~\cite{kane2014topological} of the main text. This additional symmetry plays a similar role as time reversal symmetry in magnetic space groups. It can be combined with any of the above four unitary symmetries. Hence there are eight elements to the point group. \\

In ${\bf k}$-space, the symmetry becomes more interesting. Remarkably, four of the eight elements of the preceding symmetry group transform a given point in ${\bf k}$ space into itself. This is because $C_2$ sends ${\bf k} \to -{\bf k}$ and so does $K$. The symmetry group at a ${\bf k}$ point is then
\begin{equation}
G(k) = \{e,KC_2^{site}({\bf k}), KMC_2^{hex}, MT_1^{1/2}\}, \nonumber
\end{equation}
which is again $Z_2\times Z_2$. \\

The consequences of this symmetry are the following:

\begin{enumerate}
\item The physical frequencies have doubly degenerate bands and a line node must have double degeneracy.
\item The rigidity matrix ${\mathcal R}({\bf k})$ can be made real and placed in a block diagonal form with two blocks ${\mathcal R}_+({\bf k})$ and ${\mathcal R}_-({\bf k})$, one for $MT_1^{1/2} = +1$ and one for $MT_1^{1/2}=-1$.
\item Either $\eta_+ \equiv \text{sign}\ \text{Det} [{\mathcal R}_+({\bf k})]$ or $\eta_- \equiv \text{sign}\ \text{Det} [{\mathcal R}_-({\bf k})]$ serves as a topological invariant that demands the gap closes if it changes. They must change sign at the Dirac line nodes.
\item The previous topology characterized by the topological invariant $\eta$ is rendered trivial for $\eta=\eta_+\eta_-$ never changes sign in the Brillouin zone and so never changes so long as this symmetry is preserved.
\end{enumerate}

The remaining parts of this section prove these consequences.\\

To prove that the physical frequencies are doubly degenerate we start with the following observation. If the mirror transformation $M$ in the spin space is chosen to be the $xy$-plane, then when acting on the spin deviations ${\bf x}({\bf k})$ it takes the form of a $\tau_z$ Pauli matrix. Since the Poisson bracket tensor in this space is $\sigma = \iota\tau_y$ we discover
\begin{equation}
  \{KMC_2^{hex}, {\bf{\sigma}} {\mathcal R}^T(-{\bf{k}}) {\mathcal R}({\bf{k}})\} = 0~;~
  \{MT_1^{1/2}, {\bf{\sigma}} {\mathcal R}^T(-{\bf{k}}) {\mathcal R}({\bf{k}})\} = 0. \nonumber
\end{equation}
Namely, transformations involving the mirror symmetry $M$ in the spin space anti-commute with the matrix entering the eigenvalue problem for the physical frequencies. This in turn implies the eigenvalues of this matrix come in pairs with opposite signs since if $v$ is an eigenvector of ${\bf{\sigma}} {\mathcal R}^T(-{\bf{k}}) {\mathcal R}({\bf{k}})$ with eigenvalue $\iota\omega$ then $MT_1^{1/2}\cdot v$ is an eigenvector with eigenvalue $-\iota\omega$. But the physical frequency this eigenmode corresponds to is still the same positive number $\omega$. Hence the bands are doubly degenerate.\\

To prove that the rigidity matrix takes a block diagonal form, we need to transform to the basis where the representation of the group that acts on ${\mathcal R}({\bf{k}})$ in k-space is irreducible.
We carry this out in two steps. First we diagonalize matrices corresponding to the $MT_1^{1/2}$ transformation. Since the square of this symmetry is the identity, it has eigenvalues $\pm 1$ as noted above. Thus in the basis that diagonalizes this matrix, ${\mathcal R}({\bf{k}})$ forms two blocks. Next we transform from this basis to a new basis which diagonalizes the antiunitary transformation $KC_2^{site}({\bf k})$. Since the square of this symmetry is also the identity, we can transform to a basis where ${\mathcal R}({\bf{k}})$ is real. Because the symmetry is antiunitary, however, we can find a complete basis where all eigenvectors have eigenvalue $KC_2^{site}({\bf k})=+1$.
We do so as follows. Starting with one eigenvector $v_1$ of $MT_1^{1/2}$ we create $w_1 = v_1 + KC_2^{site}({\bf k})\cdot v_1$ if this is non-zero otherwise we set $w_1 = \iota v_1$. We then take another eigenvector $v_2$ and create $w_2$ in a similar way. But we further apply Gram-Schmidt to $w_2$ following Wigner~\cite{wigner1960normal} to ensure it is orthogonal to $w_1$. We then continue with each eigenvector $v$ in $MT_1^{1/2}$. In the end, the procedure produces a basis $\{w_i\}$ which has eigenvalues $MT_1^{1/2}=\pm 1$ and $KC_2^{site}({\bf k})=1$. Since there are no further symmetries to use because $KMC_2^{hex}$ is just the product of these two symmetries, this is the eigenbasis of the symmetry group at a ${\bf k}$ point. \\

To define a continuous basis throughout the Brillouin zone, one needs to adopt a smooth gauge of parallel transport by comparing this basis to the one similarly produced at a nearby k-point ${\bf k}'$. The procedure involves the non-abelian form of the Berry connection, specifically the overlap matrix between the two bases at adjacent points ${\bf k}$ and ${\bf k}'$ as $\mathcal{B}_{ij} = v_i^\dagger({\bf k})\cdot v_j({\bf k}')$   ~\cite{soluyanov2012smooth}. One then computes the singular value decomposition $\mathcal{B}= U\cdot\Sigma\cdot V^\dagger$ and carries out a unitary transformation of the basis at ${\bf k}'$ as
\begin{equation}
 v_i ({\bf k}') = \sum_j (V\cdot U^\dagger)_{ji}  ~ v_j ({\bf k}') \nonumber
\end{equation}
This is done separately for the eigenvalue $MT_1^{1/2}= 1$ and $MT_1^{1/2}=-1$ bases. The resulting matrix $\mathcal{B}' = U\cdot\Sigma\cdot U^\dagger$ is Hermitian and thus defines a parallel transport of the vectors and a continuous basis in the Brillouin zone as required. \\

Finally, the antisymmetry and the block-diagonal form of ${\mathcal R}$ additionally explains the double degeneracy of the Dirac line nodes, the behavior of the new topological invariants $\eta_+$, $\eta_-$ and the behavior of the old topological invariant $\eta$. Consider an eigenvector $v$ of ${\bf{\sigma}} {\mathcal R}^T(-{\bf{k}}) {\mathcal R}({\bf{k}})$ with eigenvalue $\iota\omega$. The eigenvector $MT_1^{1/2}\cdot v$ has eigenvalue $-\iota\omega$ by the antisymmetry. This further implies in this subspace of two vectors that the transformation $MT_1^{1/2} = \sigma_x$, where $\sigma_x$ is the usual Pauli matrix. But this matrix has two eigenvalues $\pm1$ and so in the basis that places ${\mathcal R}({\bf{k}})$ in block diagonal form and makes it real, one zero mode belongs to the $MT_1^{1/2}= 1$ sector and one belongs to the $MT_1^{1/2}=- 1$ sector. Hence, as $\omega$ passes through zero at a line node, both $\eta_+$ and $\eta_-$ change sign and their product $\eta$ always remains the same throughout the Brillouin zone.

\bibliographystyle{unsrt}
\bibliography{reference_ordered}